\documentclass[a4paper]{article}

\usepackage{graphicx}
\usepackage{mathtools}
\usepackage{multirow}
\usepackage[utf8]{inputenc}
\usepackage{amsfonts}
\usepackage[margin=1.2in]{geometry}

\providecommand{\keywords}[1]
{
  \small	
  \textbf{\textit{Keywords---}} #1
}

\usepackage[sorting=none]{biblatex}
\begin{document}

\title{Elastic Stress Field beneath a Sticking Circular Contact under Tangential Load}

\author{E. Willert \\
Technische Universit\"at Berlin\\
Stra\ss{}e des 17. Juni 135, 10623 Berlin, Germany \\
e.willert@tu-berlin.de}
\maketitle

\begin{abstract}
Based on a potential theoretical approach, the subsurface stress field is calculated for an elastic-half space, which is subject to normal and uniaxial tangential surface tractions that -- in the case of elastic decoupling -- correspond to rigid normal and tangential translations of a circular surface domain. The stress fields are obtained explicitly and in closed form as the imaginary parts of compact complex-valued expressions. The stress state in the surface and on the central axis are considered in detail. As, within specific approximations that have been discussed at length in the literature, any tangential contact problem with friction can be understood as a certain incremental series of such rigid translations, the solutions presented here can serve as the basis of very fast superposition algorithms for the analysis of subsurface stress fields in general tangential contact problems with friction.
\end{abstract}


\section{Introduction}\label{sec1} 

The solution to a contact mechanical problem is often only concerned with the relations between macroscopic quantities (forces and global displacements), the size and shape of the contact domain, and the traction vector therein. The subsurface stress field beneath the contact is not often considered, at least in exact contact solutions; on the one hand, this is due to the fact that for the analysis of contacts as contributors to the dynamics of a multibody system, the knowledge about the relations between macroscopic quantities is sufficient; on the other hand, the exact analysis of the subsurface stress field is generally quite a complicated mathematical task. A fine example for actually both reasons is the classical paper of Hertz \cite{Hertz1882}. Based only on his solution for the force and indentation depth in the ("point") contact of smooth elastic bodies, he considered a dynamic multibody problem -- the impact of elastic spheres (in the quasi-static limit); on the other hand, the exact solution for the subsurface stresses he considered mathematically close to impossible, although undoubtedly acknowledging their importance \cite{Huber1904} in various circumstances, e.g., for the analysis of subsurface yield.

Since the work of Hertz, of course, significant (although still quite limited) progress has been made with respect to the exact determination of subsurface stress fields in mechanical contact problems. Theoretically speaking, at least for linearly elastic problems, the knowledge about the traction vector in the contact domain (which commonly is a central part of what is considered a "contact solution") would be sufficient for the determination of the complete stress field, based on the superposition of the fundamental solutions (including the corresponding stress fields) by Boussinesq \cite{Boussinesq1885} and Cerruti \cite{Cerruti1882} for normal and tangential point loading at the surface of an elastic half-space. However, the resulting integrals are -- with very few exceptions -- intractable in exact form and computationally costly if evaluated numerically.

A very powerful possibility lies in the application of potential theoretical methods. By this means, the solutions for the subsurface stress fields have been obtained in closed analytical form for the axisymmetric \cite{Huber1904} and elliptical \cite{SackfieldHills1983} frictionless Hertzian contact, as well as the axisymmetric (\cite{HamiltonGoodman1966}, \cite{Hamilton1983}) and elliptical \cite{SackfieldHills1983b} sliding Hertzian contact. Alternatively, at least for the axisymmetric frictionless normal contact, one can apply Hankel transforms, as pioneered by Sneddon \cite{Sneddon1965}; based on this procedure, very recently the solution for the stress field beneath an axisymmetric punch with a profile in the form of a power-law has been published \cite{Yangetal2022}, albeit in integral form.

To the author's knowledge, no other exact and explicit solutions have been published for the subsurface stress fields in elastic point contacts (for line contacts, the full stress field can be obtained easily based on Muskhelishvili's \cite{Muskhelishvili1958} potential). However, recently (\cite{Willertetal2020}, \cite{Forsbach2020}) it was suggested to calculate the subsurface stress state for elastic contacts with arbitrary axisymmetric convex profiles under normal and tangential load based on the superposition of rigid incremental translations of circular contact domains -- an ingenious idea to solve axisymmetric contact problems, which stems from Mossakovski \cite{Mossakovski1963} and later J\"ager \cite{Jaeger1995}, and which is described comprehensively in the handbook by Popov et al. \cite{Popovetal2019}. This superposition, obviously, requires the full knowledge of the solution for a single (normal or tangential) rigid translation of a circular contact domain (for elastically decoupled problems, this corresponds to simple axisymmetric distributions of the normal or tangential contact stress, respectively). While the authors in \cite{Willertetal2020} argue, that the respective subsurface stress fields can be obtained as certain derivatives of the corresponding known solutions for a parabolic (i.e., Hertzian) contact, they fail to give the explicit solutions for the full stress fields they require for their superposition procedure (the aforementioned derivatives are extremely lengthy and impractical). Therefore, in the present manuscript, these stress fields shall be given in explicit closed form, based on a potential theoretical approach.

\section{Problem Statement}\label{sec2}

Let us consider a linearly elastic body that obeys the restrictions of the half-space approximation; the elastic material shall have the shear modulus $G$ and Poisson's ratio $\nu$ and occupy the half-space $z \geq 0$ in a cartesian coordinate system $\left\lbrace x,y,z \right\rbrace$. Let there be loading in the form of normal and tangential tractions on a circular region with radius $a$ of the boundary surface $z = 0$ of the half-space. We will consider the following two loading scenarios: Firstly, there shall be loading in the form of a normal compressive stress according to the boundary conditions
\begin{align}
\sigma_{yz}\left(z=0\right) &= \sigma_{xz}\left(z=0\right) = 0, \nonumber \\
\sigma_{zz}\left(z=0\right) &= -\frac{p_0 a}{\sqrt{a^2 - r^2}} \quad , \quad r < a.
\label{eq_BQ_normal}
\end{align}
Here, $\sigma_{ij}$ with $i,j = \left\lbrace x,y,z \right\rbrace$ are the components of the stress tensor, $r = \sqrt{x^2 + y^2}$ is the polar radius, and $p_0$ is a constant. On the other hand, let us analyze the case of uniaxial tangential tractions in the form
\begin{align}
\sigma_{yz}\left(z=0\right) &= \sigma_{zz}\left(z=0\right) = 0, \nonumber \\
\sigma_{xz}\left(z=0\right) &= -\frac{\tau_0 a}{\sqrt{a^2 - r^2}} \quad , \quad r < a,
\label{eq_BQ_tangential}
\end{align}
with a constant $\tau_0$.

While that is not too relevant for the solution to the thus stated  problem in linear elasticity, to give some physical background, we shall briefly discuss, how the boundary conditions \eqref{eq_BQ_normal} and \eqref{eq_BQ_tangential} could be realized in a mechanical contact problem.

If the elastic half-space is incompressible (i.e., $\nu = 0.5$), the boundary conditions are easily implemented by rigid translations of the circular contact domain ($z = 0 \land r \leq a$) in the $z$- and $x$-direction, respectively; in other words, by normal and tangential loading with a rigid cylindrical flat punch. However, for a compressible material, there will be elastic coupling between the normal and tangential contact problems, and the physical realization of the boundary conditions  \eqref{eq_BQ_normal} and \eqref{eq_BQ_tangential} will not be as straightforward as for the incompressible case. Nonetheless, a common framework of analytical contact mechanics to solve contact problems with friction is the theory of Cattaneo \cite{Cattaneo1938} and Mindlin \cite{Mindlin1949}, which neglects (among other details) elastic coupling contributions. In \cite{Munisamyetal1994}, the Cattaneo-Mindlin approximate theory was compared to a rigorous numerical contact solution for the frictional Hertzian contact under shear load, and it was found, that the error of the approximation (in terms of, e.g., contact tractions) is generally small. Accordingly, the elasticity problems formulated in the above equations \eqref{eq_BQ_normal} and \eqref{eq_BQ_tangential} are a fundamental basis for the analysis of any tangential contact problem with friction of axisymmetric elastic bodies, as has been discussed in much detail in the handbook \cite{Popovetal2019}. 

\section{Solution for the Normal Load}\label{sec3}

Let us start with the solution to the boundary value problem characterized by the boundary conditions \eqref{eq_BQ_normal}. As the corresponding potential theoretical problem has been solved, and only the stress field needs to be computed from the known potential, this section can be kept short.

\subsection{Potential Solution}

According to Barber (\cite{Barber2018}, pp. 64~f.), the stress and displacement fields in the elastic half-space can be expressed in terms of the axisymmetric harmonic potential
\begin{align}
\varphi\left(r,z\right) &= -p_0 a ~ \mathfrak{Re}\left\lbrace \int_0^a F\left(r,z;\xi \right) \text{d} \xi\right\rbrace, \nonumber \\
F(r,z;\xi) &= \ln \left(\sqrt{r^2 + \left[ z + i \xi \right]^2} + z + i \xi\right),
\label{eq_norm_potential}
\end{align}
with the imaginary unit $i = \sqrt{-1}$.

\subsection{Stress Field in Terms of the Potential}

The stress field in cylindrical coordinates $\left\lbrace r, \theta, z\right\rbrace$ can be calculated from the potential based on the general relations (\cite{Barber2018}, p. 544)
\begin{align}
\sigma_{rr} &= z \frac{\partial^3 \varphi}{\partial r^2 \partial z} + \frac{\partial^2 \varphi}{\partial r^2} - 2\nu \left(\frac{\partial^2 \varphi}{\partial r^2} + \frac{\partial^2 \varphi}{\partial z^2} \right), \nonumber \\
\sigma_{\theta \theta} &= -(1 - 2\nu)\frac{\partial^2 \varphi}{\partial r^2} - \frac{\partial^2 \varphi}{\partial z^2} - z \frac{\partial^3 \varphi}{\partial r^2 \partial z} - z \frac{\partial^3 \varphi}{\partial z^3}, \nonumber \\
\sigma_{rz} &= z \frac{\partial^3 \varphi}{\partial r \partial z^2}, \nonumber \\
\sigma_{zz} &= z \frac{\partial^3 \varphi}{\partial z^3} - \frac{\partial^2 \varphi}{\partial z^2}.
\label{eq_norm_stressfrompotential}
\end{align}
All other stress components vanish because of the problem symmetry.

\subsection{Explicit Solution for the Stress Field}

Putting the solution for the elastic potential \eqref{eq_norm_potential} into the general relations \eqref{eq_norm_stressfrompotential}, and executing the resulting derivatives and integrals does not pose mathematical difficulties. The resulting stress field is given by the imaginary parts of the complex-valued field
\begin{align}
\frac{\hat{\sigma}_{rr}(r,z)}{p_0 a} &= -\frac{z}{r^2}\frac{u\left(2r^2 + u^2\right)}{\left(r^2 + u^2\right)^{3/2}} + \frac{1}{\sqrt{r^2 + u^2}} + \frac{1-2\nu}{r^2}\left(u - \sqrt{r^2 + u^2}\right), \nonumber \\
\frac{\hat{\sigma}_{rz}(r,z)}{p_0 a} &= \frac{rz}{\left(r^2 + u^2\right)^{3/2}}, \nonumber \\
\frac{\hat{\sigma}_{zz}(r,z)}{p_0 a} &= \frac{uz}{\left(r^2 + u^2\right)^{3/2}} +  \frac{1}{\sqrt{r^2 + u^2}}, \nonumber \\
\frac{\hat{\sigma}_{\theta \theta}(r,z)}{p_0 a} &= \frac{2\nu}{\sqrt{r^2 + u^2}} + \frac{uz}{r^2\sqrt{r^2 + u^2}} - \frac{1-2\nu}{r^2}\left(u - \sqrt{r^2 + u^2}\right),
\label{eq_norm_complexstress}
\end{align}
with the complex auxiliary variable
\begin{align}
u = z + ia.
\label{eq_def_u}
\end{align}

\subsection{Stresses in the Surface and on the Axis of Symmetry}

In the surface $z = 0$, the corresponding imaginary parts of the complex-valued field \eqref{eq_norm_complexstress} can be explicitly evaluated easily. We obtain for the stress state in the surface inside the contact domain
\begin{align}
\frac{\sigma_{rr}(r < a,z=0)}{p_0 a} &= -\frac{1}{\sqrt{a^2 - r^2}} + \frac{1-2\nu}{r^2}\left(a - \sqrt{a^2 - r^2}\right), \nonumber \\
\frac{\sigma_{zz}(r < a,z=0)}{p_0 a} &= -\frac{1}{\sqrt{a^2 - r^2}}, \nonumber \\
\frac{\sigma_{\theta \theta}(r < a,z=0)}{p_0 a} &= -\frac{2\nu}{\sqrt{a^2 - r^2}} - \frac{1-2\nu}{r^2}\left(a - \sqrt{a^2 - r^2}\right),
\label{eq_norm_stress_surface_in}
\end{align}
and outside the contact domain
\begin{align}
\frac{\sigma_{rr}(r > a,z=0)}{p_0 a} = -\frac{\sigma_{\theta \theta}(r > a,z=0)}{p_0 a} = \frac{\left(1 - 2\nu \right)a}{r^2}.
\label{eq_norm_stress_surface_out}
\end{align}
All other stress components vanish. The results \eqref{eq_norm_stress_surface_in} and \eqref{eq_norm_stress_surface_out} are in perfect agreement with the ones reported in \cite{Willertetal2020} for the same problem.

On the axis of symmetry, $r = 0$, the stresses have to be reworked. For the non-vanishing physical stresses one obtains
\begin{align}
\frac{\sigma_{rr}(r = 0,z)}{p_0 a} &= \frac{\sigma_{\theta \theta}(r = 0,z)}{p_0 a} = -\frac{a^3}{\left(a^2 + z^2\right)^2} + (1 - 2\nu)\frac{a}{2\left(a^2 + z^2\right)}, \nonumber \\
\frac{\sigma_{zz}(r = 0,z)}{p_0 a} &= -\frac{a^3 + 3az^2}{\left(a^2 + z^2\right)^2}.
\label{eq_norm_stress_axi}
\end{align} 

\subsection{Stress Field in Cartesian Coordinates}

While above formulation of the stress field in axisymmetric cylindrical coordinates is very compact, it sometimes may be preferable to have the field in the original cartesian coordinate system.

Based on the transformation rules for the stress tensor between cartesian and axisymmetric cylindrical coordinates,
\begin{align}
\sigma_{xx} &= \frac{x^2}{r^2}\sigma_{rr} + \frac{y^2}{r^2}\sigma_{\theta \theta} \quad , \quad
\sigma_{yy} = \frac{y^2}{r^2}\sigma_{rr} + \frac{x^2}{r^2}\sigma_{\theta \theta}, \nonumber \\
\sigma_{xy} &= \frac{xy}{r^2}\left(\sigma_{rr} - \sigma_{\theta \theta}\right) \quad , \quad
\sigma_{xz} = \frac{x}{r}\sigma_{rz} \quad , \quad \sigma_{yz} = \frac{y}{r}\sigma_{rz},
\end{align}
we obtain the stress field as the imaginary parts of the complex-valued field\footnote{$\sigma_{zz}$, of course, remains unchanged under the coordinate transform, and will therefore not be written out again}
\begin{align}
\frac{\hat{\sigma}_{xx}}{p_0 a} &= \frac{uz}{r^4}\left[\frac{y^2}{\sqrt{r^2+u^2}} - \frac{x^2\left(2r^2 + u^2\right)}{\left(r^2 + u^2\right)^{3/2}}\right] + \frac{x^2 + 2\nu y^2}{r^2 \sqrt{r^2 + u^2}} + \frac{x^2 - y^2}{r^4}(1 - 2\nu)\left(u - \sqrt{r^2 + u^2}\right), \nonumber \\
\frac{\hat{\sigma}_{yy}}{p_0 a} &= \frac{uz}{r^4}\left[\frac{x^2}{\sqrt{r^2+u^2}} - \frac{y^2\left(2r^2 + u^2\right)}{\left(r^2 + u^2\right)^{3/2}}\right] + \frac{y^2 + 2\nu x^2}{r^2 \sqrt{r^2 + u^2}} + \frac{y^2 - x^2}{r^4}(1 - 2\nu)\left(u - \sqrt{r^2 + u^2}\right), \nonumber \\
\frac{\hat{\sigma}_{xy}}{p_0 a} &= \frac{xy}{r^2}\left\lbrace-\frac{uz}{r^2}\frac{3r^2+2u^2}{\left(r^2 + u^2\right)^{3/2}} + (1-2\nu)\left[\frac{1}{\sqrt{r^2 + u^2}} + \frac{2}{r^2}\left(u - \sqrt{r^2 + u^2}\right)\right]\right\rbrace, \nonumber \\
\frac{\hat{\sigma}_{xz}}{p_0 a} &= \frac{xz}{\left(r^2 + u^2\right)^{3/2}}, \nonumber \\
\frac{\hat{\sigma}_{yz}}{p_0 a} &= \frac{yz}{\left(r^2 + u^2\right)^{3/2}}.
\end{align}

In the surface inside the contact domain, the physical stresses simplify to
\begin{align}
\frac{\sigma_{xx}(r < a,z=0)}{p_0 a} &= -\frac{x^2 + 2\nu y^2}{r^2 \sqrt{a^2 - r^2}} + \frac{x^2-y^2}{r^4}(1 - 2\nu)\left(a - \sqrt{a^2 - r^2}\right), \nonumber \\
\frac{\sigma_{yy}(r < a,z=0)}{p_0 a} &= -\frac{y^2 + 2\nu x^2}{r^2 \sqrt{a^2 - r^2}} + \frac{y^2-x^2}{r^4}(1 - 2\nu)\left(a - \sqrt{a^2 - r^2}\right), \nonumber \\
\frac{\sigma_{xy}(r < a,z=0)}{p_0 a} &= \frac{xy}{r^2}(1 - 2\nu)\left[-\frac{1}{\sqrt{a^2-r^2}} + \frac{2}{r^2}\left(a - \sqrt{a^2 - r^2}\right)\right],
\end{align}
while in the surface, but outside the contact domain the non-vanishing stress components read
\begin{align}
\frac{\sigma_{xx}(r > a,z=0)}{p_0 a} = \frac{\left(x^2-y^2\right)a}{r^4}(1 - 2\nu), \nonumber \\ 
\frac{\sigma_{yy}(r > a,z=0)}{p_0 a} = \frac{\left(y^2-x^2\right)a}{r^4}(1 - 2\nu), \nonumber \\
\frac{\sigma_{xy}(r > a,z=0)}{p_0 a} = \frac{2xya}{r^4}(1 - 2\nu).
\end{align}

On the axis of symmetry, we obtain from Eqs. \eqref{eq_norm_stress_axi}
\begin{align}
\frac{\sigma_{xx}(r = 0,z)}{p_0 a} &= \frac{\sigma_{yy}(r = 0,z)}{p_0 a} = -\frac{a^3}{\left(a^2 + z^2\right)^2} + (1 - 2\nu)\frac{a}{2\left(a^2 + z^2\right)}.
\end{align}

\section{Solution for the Tangential Load}\label{sec4}

We now turn our attention to the boundary value problem characterized by the boundary conditions \eqref{eq_BQ_tangential}.

\subsection{Potential Solution}

Applying the potential theoretical procedure suggested by Hamilton \& Goodman \cite{HamiltonGoodman1966} (for the sliding circular Hertzian contact) to the boundary conditions \eqref{eq_BQ_tangential}\footnote{There seems to be missing a minus sign in Eq. (5) of \cite{HamiltonGoodman1966}}, one quickly determines that the displacement and stress fields in the elastic half-space, under the action of the tangential loading \eqref{eq_BQ_tangential}, can be expressed in terms of the harmonic potential
\begin{align}
T\left(r,u\right) = \tau_0 a ~ \mathfrak{Im}\left\lbrace \frac{1}{2}\left(u^2 - \frac{r^2}{2}\right)\ln\left(u + \sqrt{r^2 + u^2}\right) - \frac{3}{4}u \sqrt{r^2 + u^2} + \frac{r^2}{4}\right\rbrace,
\end{align}
with the complex auxiliary variable $u$ as in Eq. \eqref{eq_def_u}. 

\subsection{Stress Field in Terms of the Potential}

According to \cite{HamiltonGoodman1966}, the displacement field can be written in terms of the potential as follows:
\begin{align}
2G u_x &= 2\nu \frac{\partial^2 T}{\partial x^2} + 2\frac{\partial^2 T}{\partial z^2} - z\frac{\partial^3 T}{\partial x^2 \partial z}, \nonumber \\
2G u_y &= 2\nu \frac{\partial^2 T}{\partial x \partial y} - z\frac{\partial^3 T}{\partial x \partial y \partial z}, \nonumber \\
2G u_z &= \left(1 - 2\nu \right) \frac{\partial^2 T}{\partial x \partial z} - z\frac{\partial^3 T}{\partial x \partial z^2}.
\end{align}
Applying Hooke's law and accounting for the fact that $T$ is harmonic, one obtains the stresses in terms of the potential,
\begin{align}
\sigma_{xy} &= 2\nu \frac{\partial^3 T}{\partial x^2 \partial y} + \frac{\partial^3 T}{\partial y \partial z^2} - z\frac{\partial^4 T}{\partial x^2 \partial y \partial z}, \nonumber \\
\sigma_{xz} &= \frac{\partial^3 T}{\partial z^3} - z\frac{\partial^4 T}{\partial x^2 \partial z^2}, \nonumber \\
\sigma_{yz} &=  - z\frac{\partial^4 T}{\partial x \partial y \partial z^2}, \nonumber \\
\sigma_{xx} &= 2\nu \frac{\partial^3 T}{\partial x^3} + 2(1 + \nu)\frac{\partial^3 T}{\partial x \partial z^2} - z\frac{\partial^4 T}{\partial x^3 \partial z}, \nonumber \\
\sigma_{yy} &= -2\nu \frac{\partial^3 T}{\partial x^3} - z\frac{\partial^4 T}{\partial x \partial y^2 \partial z}, \nonumber \\
\sigma_{zz} &= - z\frac{\partial^4 T}{\partial x \partial z^3}.
\label{eq_tang_stressfrompotential}
\end{align}

\subsection{Explicit Solution for the Stress Field}

The derivatives in Eqs. \eqref{eq_tang_stressfrompotential} can be evaluated without severe difficulties. The resulting stress field is given by the imaginary parts of the complex-valued field
\begin{align}
\frac{\hat{\sigma}_{xy}}{\tau_0 a} &= 2\nu \left[\frac{y}{r^2}\left(\frac{u}{\rho + u} - \frac{1}{2}\right) + \frac{x^2y}{\rho\left(\rho + u\right)^3}\right] + \frac{y}{\rho\left(\rho + u\right)} - z\left[\frac{y}{\rho\left(\rho + u\right)^2} - \frac{x^2y\left(3\rho + u\right)}{\rho^3 \left(\rho + u\right)^3}\right], \nonumber \\
\frac{\hat{\sigma}_{xz}}{\tau_0 a} &= \frac{1}{\rho} - z\left[\frac{1}{\rho\left(\rho + u\right)} - \frac{x^2\left(2\rho + u\right)}{\rho^3 \left(\rho + u\right)^2}\right], \nonumber \\
\frac{\hat{\sigma}_{yz}}{\tau_0 a} &= \frac{xyz\left(2\rho + u\right)}{\rho^3 \left(\rho + u\right)^2}, \nonumber \\
\frac{\hat{\sigma}_{xx}}{\tau_0 a} &= 2\nu \left[\frac{3x}{r^2}\left(\frac{u}{\rho + u} - \frac{1}{2}\right) + \frac{x^3}{\rho\left(\rho + u\right)^3}\right] + \frac{2x(1 + \nu)}{\rho\left(\rho + u\right)} - z\left[\frac{3x}{\rho\left(\rho + u\right)^2} - \frac{x^3\left(3\rho + u\right)}{\rho^3 \left(\rho + u\right)^3}\right], \nonumber \\
\frac{\hat{\sigma}_{yy}}{\tau_0 a} &= -2\nu \left[\frac{3x}{r^2}\left(\frac{u}{\rho + u} - \frac{1}{2}\right) + \frac{x^3}{\rho\left(\rho + u\right)^3}\right] - z\left[\frac{x}{\rho\left(\rho + u\right)^2} - \frac{xy^2\left(3\rho + u\right)}{\rho^3 \left(\rho + u\right)^3}\right], \nonumber \\
\frac{\hat{\sigma}_{zz}}{\tau_0 a} &= \frac{xz}{\rho^3},
\end{align}
with the complex auxiliary variable
\begin{align}
\rho = \sqrt{r^2 + u^2} = \sqrt{r^2 + z^2 - a^2 + 2iaz}.
\end{align}

\subsection{Stresses in the Surface and on the Central Axis}

In the surface, inside the contact domain, the only non-vanishing physical stress component is
\begin{align}
\frac{\sigma_{xz}(r < a, z = 0)}{\tau_0 a} = -\frac{1}{\sqrt{a^2 - r^2}}.
\end{align}
In the surface, but outside the contact domain, the non-vanishing physical stress components are
\begin{align}
\frac{\sigma_{xy}(r > a, z = 0)}{\tau_0 a} &= -\frac{ya}{r^4}\left[ \frac{r^2}{\sqrt{r^2-a^2}} - 2\nu\left\lbrace \left(1 - \frac{4x^2}{r^2}\right)\sqrt{r^2-a^2} + \frac{x^2}{\sqrt{r^2-a^2}}\right\rbrace \right], \nonumber \\
\frac{\sigma_{xx}(r > a, z = 0)}{\tau_0 a} &= -\frac{xa}{r^4}\left[ \frac{2r^2}{\sqrt{r^2-a^2}} - 2\nu\left\lbrace \left(3 - \frac{4x^2}{r^2}\right)\sqrt{r^2-a^2} - \frac{y^2}{\sqrt{r^2-a^2}}\right\rbrace \right], \nonumber \\
\frac{\sigma_{yy}(r > a, z = 0)}{\tau_0 a} &= -2\nu \frac{xa}{r^4}\left[\left(3 - \frac{4x^2}{r^2}\right)\sqrt{r^2-a^2} + \frac{x^2}{\sqrt{r^2-a^2}} \right],
\end{align}
which agrees with the results reported in \cite{Willertetal2020} for the same problem.

On the $z$-axis, the only non-vanishing physical stress component is
\begin{align}
\frac{\sigma_{xz}(r = 0,z)}{\tau_0 a} = -\frac{a^3}{\left(a^2+z^2\right)^2}.
\end{align}

\section{Conclusions}\label{sec5}

Based on known potential theoretical procedures and solutions, the subsurface stress fields have been calculated exactly for an elastic half-space, which is subject to surface tractions that -- in the case of elastic decoupling -- correspond to rigid normal and tangential translations of a circular surface domain. Within the framework of the Cattaneo-Mindlin approximation, any tangential frictional contact problem of axisymmetric elastic bodies can be solved as a specific series of such (incremental) rigid translations \cite{Popovetal2019}. In this sense, the presented solutions allow for a very fast calculation of subsurface stress fields for arbitrary axisymmetric elastic tangential contacts with friction.

It should be noted that the obtained solutions are mathematically exact within the restrictions of the physical modelling (namely the assumptions of linear elasticity and the validity of the half-space approximation), and can thus also serve as benchmark solutions for different numerical contact algorithms.

\section{Acknowledgements}

This work was supported by the German Research Foundation under the project number PO 810/66-1.

\printbibliography

\end{document}